# Quantum-dot single-photon sources for the quantum internet

High-performance quantum light sources based on semiconductor quantum dots coupled to microcavities are showing their promise in long-distance solid-state quantum networks.

Chao-Yang Lu and Jian-Wei Pan

By harnessing the laws of quantum mechanics, quantum internet[1] is expected to expand the capabilities of the classical internet. The strange phenomena in the quantum world, such as superposition and entanglement, are recognized highly non-classical resources that enable information processing tasks otherwise impossible with classical means. Functional elements of a future quantum internet (see Fig. 1) will include quantum key distribution[2], which allows distant users to share a secret key for secure transmissions of sensitive classical information. Quantum internets will also provide a way to transfer quantum information from one node to another over long distances, based on a protocol called quantum teleportation[3]. Teleportation is not only useful in extending the quantum communication distances, but also essential in distributed quantum computation. Other functional components of the quantum internet include distributions of entanglement over nodes[4], secure access to cloud quantum computers[5], distributed quantum sensing[6], quantum storage[7], and integration of ground-space quantum channels using satellites[8] (see Fig. 1).

Photons, the fast-flying qubits, are most suitable for transferring information over long distance. An important element for the quantum internet is the quantum light sources. The Bennett-Brassard-1984 quantum key distribution protocol[2] requires an ideal single-photon source that emits one and only one photon each time. To obtain a high secret key rate after long-distance transmission, it is desirable that the single-photon sources have high efficiency, high repetition rate, and at telecommunication wavelength. More advanced quantum networking tasks, such as quantum teleportation and entanglement swapping[9], further require the independent photons to be Fourier-transform-limited to allow quantum interference.

Building blocks of quantum networks have been demonstrated using nonlinear optics[10], trapped atoms[11], color centers in diamond[12], etc. Recent advances in nanotechnologies, particularly, in the optically-active quantum dots (QDs) embedded in microcavities[13], are expected to open up new routes to solid-state quantum networks. This Commentary mainly focuses on epitaxial grown single QDs[14]. Among different types of quantum emitters, the QDs offer unmatched advantages, such as near-unity quantum efficiency, large dipole moment, narrow optical transition lines, monolithically grown nanocavities, interface with spin qubits, and an on-chip platform. The overall high performance[13] of the epitaxial QDs has established a benchmark for other emerging solid-state emitters to compare.

A single QD can behave as a single, isolated artificial atom. Therefore, the second-order correlation of its radiative emission is anti-bunched, i.e., there are never two photons emitted at the same time. The emitted single photons are indistinguishable if the individual photons are identical in all degrees of freedom including polarization, spatial, spectral, and temporal modes. The QD photon indistinguishability benefits from low-defect-density material growth and resonant excitation. Resonant excitation[15] directly addresses the two-level system in the solid state, thus avoiding the time jitter associated with non-resonant excitation. The QD excited state is deterministically prepared at $\pi$ pulse, which requires laser power typically orders of magnitude lower than that in non-resonant excitation,thus greatly reducing the excitation-induced dephasing.

The single-photon source efficiency—defined as the percentage of the output single-photon counts per second before arriving at the detectors to the pump repetition rate—depends on the quantum efficiency, excited-state preparation efficiency, extraction efficiency, and the optical path transmission efficiency. The extraction efficiency can be enhanced using Purcell microcavities, where the dipole emission is funnelled into a guided mode. By coupling the single QDs to microcavities[16-18] with optimized spatial and spectral overlap, the pulsed resonance fluorescence single-photon source efficiency has reached 57% in open-microcavity[19]. Due to Purcell acceleration, the spontaneous decay lifetime has been shortened to ~50 ps in microcavities[19] from the typical ~1 ns in bulk GaAs, which indicates that the pumping repetition rate can be increased to tens of GHz. In addition, the shortening of the lifetime further mitigated the dephasing of the QD, enabling a 98% photon indistinguishability for independent single photons emitted with a 1.5-µs time separation[19].

The simultaneous combination of the efficiency, purity, and indistinguishability in a QD device[18,19] is getting closer to a perfect quantum light source that the community has been dreaming for more than 20 years. One immediate application is optical quantum computing that uses single photons as elementary qubits. An intermediate, non-universal model of liner optical quantum computing is boson sampling[20], which requires only high-quality single-photon source, linear optical network, and photon detection. Boson sampling significantly reduces the experimental demands—compared to Shor's algorithm—but can provide even more compelling complexity-theoretic evidence for the quantum computational speedup conjectured by Feynman in 1981. Indeed, boson sampling[20] has been proposed as an important candidate to demonstrate quantum computational advantage—a task that no classical computer can solve within a reasonable amount of time.

Early proof-of-principle demonstration of boson sampling was performed in 2013 with 3 pseudo-single photons from parametric down-conversion, a non-linear optical process where, if a strong laser beam is focused on a nonlinear crystal, the pump photons have some tiny probability to split into correlated pairs of lower energy. The single-photon sources based on the QDs in micropillars[16,18], continuously developed with increasingly higher efficiency, purity, and indistinguishability, have scaled up the boson sampling to 5 photons[21] in 2017 and 20 photons[22] in 2019. These multi-photon experiments rely on simultaneous detection of $N$ photons, with an overall probability scales as $p^N$ where $p$ is the efficiency of the single-photon source. Therefore, the 20-photon detection rate using the state-of-the-art QD-based source[19] will be 35 orders of magnitude higher than using the best parametric down-conversion source[23]. These multi-photon experiments are powerful examples that have showed a superiority of the QD-based single photons compared to the probabilistic sources for scaling up photonic quantum technologies.

Similarly, the high-performance quantum light sources are also expected to enhance the capability of quantum communications. The main task for the quantum key distribution to become practical is to go to long distances and ensure security with realistic devices[24]. However, most of the quantum communications based on QDs so far were limited to short distances[25-27]. One of the challenge is the wavelength of the QD emission. So far, the brightest single-photon sources are based on InAs QDs which emit at a wavelength range of 850-1000 nm, which need to be frequency converted to ~1550 nm to exploit the low transmission loss in telecommunication fibers. Meanwhile, the QDs directly emitting single photons in the telecommunication band have not reached an overall high performance comparable to the InAs QDs yet, and are under ongoing development.

For quantum teleportation and entanglement swapping between remote QDs, besides the indistinguishable single photons, remaining challenges include frequency matching, time synchronization, and high-fidelity transmission over long fibers in the field. The wavelength of different QDs can be tuned to be the same using gate voltage or strain. Another more "non-invasive" method is to compensate the energy difference by tuning the pump laser in the frequency conversion (to ~1550 nm), killing two birds with one stone[28]. To synchronize the two independent photons, each from a fluctuating channel, to arrive simultaneously at a middle point for quantum interference, active stabilization strategies are required to compensate the fluctuation of their arrival time. Using a closed feedback loop[29], the time jitter between independent photons separated by 50-km fibers has been reduced to ~7 ps. Which such a time jitter is much larger than the coherence time (~100 fs) of typical unfiltered down-conversion sources, it is already much shorter than typical coherence time of the QD photons (~100 ps).

The quantum teleportation and entanglement swapping based on QDs will be important ingredients for quantum networks. They can be directly used in the quantum relay[30], a scheme based on teleportation-based quantum non-demolition measurement of photons to moderately extend the quantum communication distance by increasing the signal-to-noise ratio. Indeed, the teleportation and swapping can be seen as a three- and four-node quantum relay, respectively[30]. The functionalities and communication range of the quantum relay can be further improved by combining with suitable quantum memories to form a full-version of quantum repeaters[31].

Another crucial advantage of the QDs, compared to parametric down-conversion, is the deterministic single-photon emission and intrinsic higher-order suppression, which can allow the realization of multi-photon interferometry[32] in a non-post-selection way. Four single photons from the QDs can generate a two-photon entangled pair in an event-ready manner. The QDs can also directly emit two-photon entanglement via biexciton-exciton cascade decay. Finally, larger Greenberger-Horne-Zeilinger photonic states can be created at distant locations using "fusion"[33] of two-photon pairs, which will be useful for all-photonic quantum repeaters[34], distributed quantum sensing[6], and measurement-based quantum computation[33].

To sum up, semiconductor QDs are appealing for near-future applications ranging from high-rate quantum key distribution, long-distance quantum teleportation, multi-photon entanglement distribution across networks[35], and intermediate-scale quantum computating[36]. To this end, further methods need to be developed, such as the control of frequency, time and polarization during the faithful transmissions of the single photons over a long distance outside the laboratories. Eventually, no matter what physical system is first proven successful to build practical quantum computers in the future, to link them together as a quantum internet, photons will serve as an ideal messenger.

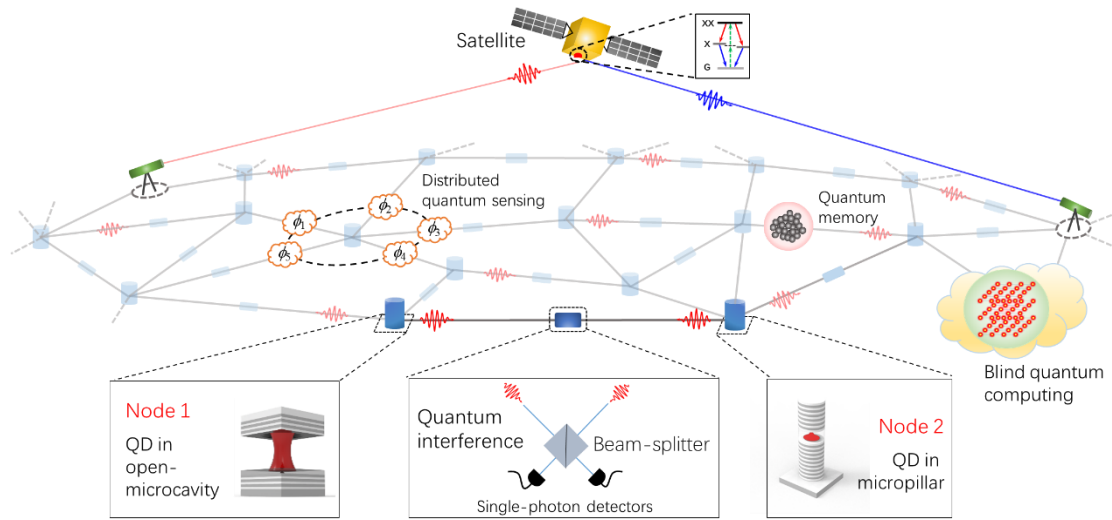

**Figure 1**: An envisioned quantum internet with semiconductor quantum dots (QDs). The nodes consist of QDs embedded in microcavities, such as micropillars, bullseyes, and open micro-cavities. High-efficiency and indistinguishable single photons from remote QDs are sent over telecommunications fibers and quantum interfered in the middle of the channel. Deterministic QD-based two-photon entanglement source can be equipped in a satellite to distribute entangled photons over a range of thousands of kilometers. The fibre-based link can be further combined with satellites to establish a space-ground integrated quantum network. The quantum internet distributes multi-photon entanglement which serves a valuable resource in quantum teleportation, entanglement swapping, distributed quantum sensing, and measurement-based quantum computation. Hybrid quantum memories of the narrowband single photons can be achieved using, for example, cold atoms or rare-earth-doped materials. The single photons can be interfaced with quantum computers based on photonic qubits or neutral atom arrays or superconducting qubits. One way to securely delegate quantum computation to an untrusted server while maintaining the client's privacy on their data and algorithms is called blind quantum computing.

Chao-Yang Lu[1,2], Jian-Wei Pan[1,2]

[1] Hefei National Laboratory for Physical Sciences at Microscale and Department of Modern Physics, University of Science and Technology of China, Hefei, Anhui 230026, China.

[2] CAS Centre for Excellence in Quantum Information and Quantum Physics, University of Science and Technology of China, Shanghai, 201315, China.